\documentstyle[12pt]{article}
\makeatletter
\@addtoreset{equation}{subsection}
\makeatother

\begin{document}
\begin{center}
{\Large \bf
Indeterministic Quantum Gravity and Cosmology} \\[0.5cm]
{\large\bf VII. Dynamical Passage through Singularities:
Black Hole and Naked Singularity, Big Crunch and
Big Bang} \\[1.5cm]
{\bf Vladimir S.~MASHKEVICH}\footnote {E-mail:
mash@phys.unit.no}  \\[1.4cm]
{\it Institute of Physics, National academy
of sciences of Ukraine \\
252028 Kiev, Ukraine} \\[1.4cm]
\vskip 1cm

{\large \bf Abstract}
\end{center}

This paper is a continuation of the papers [1-6]. The aim of
the paper is to incorporate singularities---both local (black
hole and naked singularity) and global (big bang and big crunch)
---into the dynamics of indeterministic quantum gravity and
cosmology. The question is whether a singularity is dynamically
passable, i.e., whether a dynamical process which ends with a
singularity may be extended beyond the latter. The answer is yes.
A local singularity is trivially passable, while the
passableness for a global singularity may invoke $CPT$
transformation. The passableness of the singularities implies
pulsating black holes and the oscillating universe. For the
local singularity, the escape effect takes place: In a vicinity
of the singularity, quantum matter leaves the gravitational
potential well.

\newpage

\hspace*{6 cm}
\begin{minipage} [b] {9 cm}
... put thyself into the trick of singularity.
\end{minipage}
\begin{flushright}
William Shakespeare \vspace*{0.8 cm}
\end{flushright}

\begin{flushleft}
\hspace*{0.5 cm} {\Large \bf Introduction}
\end{flushleft}

In literature, there is a great body of information on
singularities in gravity and cosmology, so that we shall
restrict our consideration of them to the theory being
developed in this series of papers. To wit, the aim of the
present paper is to incorporate the singularities into the
dynamics of indeterministic quantum gravity and cosmology
(IQGC). What is at issue is the question whether a singularity
is dynamically passable, i.e., whether a dynamical process
which ends with the singularity may be extended beyond the
latter. The answer is yes.

There are local singularities (black hole and naked singularity)
and global ones (big crunch and big bang). For any singularity,
dynamical passableness means that the singularity exists only
instantly in time, so that a gravitational contraction resulting
in the singularity is immediately followed by an expansion. The
expansion in its turn is followed by the next contraction and
so on, so that we have either a pulsating local object or the
oscillating universe.

A singularity manifests itself both in a static aspect and in a
dynamical one. The former is that the metric $g$ relating to the
singularity is degenerate, the latter is that $g$ as a function
of time cannot be $C^\infty$. It is the dynamical aspect that we
are mainly interested in. Indeed $g$ may be $C^1$ or $C$.

There are singularities of two types: tempered and strong ones.
A tempered singularity is $C^1$, and in this case a state
vector $\Psi$ of matter and energy $\varepsilon$ exist. A strong
singularity is $C$, and is that case $\Psi$ and $\varepsilon$
do not exist $(\varepsilon=\infty)$. The tempered singularity
is trivially passable. The passableness of the strong
singularity invokes $CPT$ transformation. A local singularity
is tempered. A global singularity (big crunch followed by big
bang) may be both tempered and strong, depending on the
pressure of matter.

For a local singularity, the escape effect takes place: In a
vicinity of the singularity, quantum matter leaves the
gravitational potential well before the well results in the
singularity. The effect is impossible in (semiclassical)
general relativity because of the constraint equation
(which does not hold in IQGC).

\section{Singularities and passableness}

\subsection{Future-singular point}

Let ${\cal P}_{(-\theta,0)}$ be a deterministic process (from
here on see [5]) and $g_t,\Psi_t,\varepsilon_t,\:
t\!\in\!(-\theta,
0)$, be determined by $\cal P$. The point $t=0$ is a
future-singular point for the process iff $g_{-0}\equiv
\lim_{t\to-0}g_t$ exists and is degenerate, $\dot g_{-0}
\equiv(\partial g/\partial t)_{-0}$ exists, and a finite
$\ddot g_{-0}$ does not exists. The point $t=0$ is
temperately singular iff $\Psi_{-0}$ and $\varepsilon_{-0}$
exist for a solution to
\begin{equation}
H_t\Psi_t=\varepsilon_t\Psi_t;
\label{1.1}
\end{equation}
the point $t=0$ is strongly singular iff $\Psi_{-0}$ and
$\varepsilon_{-0}$ do not exist $(\varepsilon_{-0}=\infty)$.

\subsection{Past-singular point}

A past-singular point $t=0$ is defined straigtforwardly
by the replacements
\begin{equation}
{\cal P}_{(-\theta,0)}\to{\cal P}_{(0,\theta)},\qquad -0\to +0.
\label{1.2}
\end{equation}

\subsection{Passable singular point}

Let ${\cal P}_{(-\theta^-,0)}$ be a deterministic process and
$t=0$ be a future-singular point for it. The point is passable
iff there exists a deterministic process ${\cal P}_{(0,
\theta^+)}$, such that $t=0$ is a past-singular point for it
and:

for a tempered singularity
\begin{equation}
\omega_{+0}=\lim_{t\to+0}{\cal P}_{(0,\theta^+)}(t)=
\lim_{t\to-0}{\cal P}_{(-\theta^{-},0)}(t)=\omega_{-0};
\label{1.3}
\end{equation}

for a strong singularity
\begin{equation}
\theta^+=\theta^-\equiv\theta,\qquad
{\cal P }_{(0,\theta)}(t)=CPT\:{\cal P}_{(-\theta,0)}(-t)
\label{1.4}
\end{equation}
(as to $CPT$ in IQGC see [6]).

In IQGC, for any given $\theta^+$, ${\cal P}_{(0,\theta^+)}$
should be unique.

We shall say that a passable tempered (respectively strong)
singularity is trivially (respectively $CPT$)
passable.

The passableness for a past-singular point is defined in a
similar manner.

\subsection{Local and global singularity}

For a local singular point $t=0$, the metric $g$ is degenerate
at a single point of spacetime manifold $M$. For a global
singular point $t=0$, the metric is degenerate at $t=0$ on the
unit 3-sphere, i.e., in the subset
\begin{equation}
\{0\}\times{\cal S}^3\subset M
\label{1.5}
\end{equation}
of $M$ (see [6]).

\section{Black hole and naked singularity}

\subsection{Equation of motion}

In the case of black hole or naked singularity, a singularity
is local. To solve the problem of passableness for the
singularity, it suffices to consider the simplest case: a
gravitational collapse of a spherically symmetric ball of dust
with uniform density [7,8].

The interior metric in comoving coordinates is
\begin{equation}
g=dt^2-R^{2}(t)\left[\frac{dr^{2}}{1-kr^{2}}+r^{2}d\theta^{2}
+r^{2}\sin^{2}\theta\:d\varphi^{2}\right].
\label{2.1}
\end{equation}
The singularity relates to the center of the ball, i.e.,
to $r=0$. The projected Einstein equation [2,3] gives
\begin{equation}
G_{ij}=8\pi\kappa T_{ij}\Rightarrow k+2R\ddot R+\dot R^{2}=
-8\pi\kappa pR^{2},
\label{2.2}
\end{equation}
so that for $p=0$ we have
\begin{equation}
k+2R\ddot R+\dot R^{2}=0,
\label{2.3}
\end{equation}
whence
\begin{equation}
\frac{d}{dR}[R(k+\dot R^{2})]=0,\qquad R(k+\dot R^{2})=C.
\label{2.4}
\end{equation}
The initial conditions are
\begin{equation}
R(-\tau/2)=1,\quad \dot R(-\tau/2),
\label{2.5}
\end{equation}
so that $C=k$ and we obtain
\begin{equation}
R(k+\dot R^{2})=k.
\label{2.6}
\end{equation}
Matching the interior and exterior solutions leads to
the value
\begin{equation}
k=\frac{2\kappa M}{a^{3}},
\label{2.7}
\end{equation}
where $a$ is the ball radius in the comoving coordinate system,
\begin{equation}
M=\frac{4\pi}{3}\rho(0)a^{3}
\label{2.8}
\end{equation}
is the ball mass, and $\rho(0)$ is the initial matter density.

An important point is that since the pressure on the ball
surface is always zero, the energy constancy condition
\begin{equation}
\varepsilon=M={\rm const}
\label{2.9}
\end{equation}
holds even for a nonzero internal pressure.

\subsection{Metric singularity}

The metric (\ref{2.1}) depends on
\begin{equation}
q=R^{2}
\label{2.10}
\end{equation}
rather than on $R$ itself. Thus a metric singularity should
be treated in terms of $q$. Equation (\ref{2.6}) and the
initial conditions (\ref{2.5}) take the form
\begin{equation}
\dot q^{2}+4kq-4kq^{1/2}=0,
\label{2.11}
\end{equation}
\begin{equation}
q(-\tau/2)=1,\quad \dot q(-\tau/2)=0.
\label{2.12}
\end{equation}

For $q\to 0$ we have
\begin{equation}
\dot q^{2}-4kq^{1/2}=0,
\label{2.13}
\end{equation}
so that
\begin{equation}
\frac{dq}{dt}=\mp 2k^{1/2}q^{1/4},
\label{2.14}
\end{equation}
the minus (respectively plus) sign corresponding to a
contracting (respectively expanding) ball. Choosing $\tau$ in
(\ref{2.12}) so that
\begin{equation}
q(0)=0,
\label{2.15}
\end{equation}
we obtain for $t\to 0$
\begin{equation}
q=\left(\frac{3}{2}\right)^{4/3}k^{2/3}t^{4/3},
\label{2.16}
\end{equation}
\begin{equation}
\dot q=2\left(\frac{3}{2}\right)^{1/3}k^{2/3}t^{1/3},
\label{2.17}
\end{equation}
\begin{equation}
\ddot q=\left(\frac{2}{3}\right)^{2/3}k^{2/3}t^{-2/3}.
\label{2.18}
\end{equation}

Equation (\ref{2.15}) implies that the metric (\ref{2.1}) is
degenerate, i.e., singular at $t=0$.

\subsection{Tempered singularity and trivial passableness}

Equations (\ref{2.16}), (\ref{2.17}) are valid both for
$t\le 0$ and for $t>0$; equation (\ref{2.18}) is valid
both for $t<0$ and for $t>0$. Thus equations
(\ref{2.16})-(\ref{2.18}),(\ref{2.9})
imply that the singularity
considered is a tempered one: In accordance with eq.(\ref{1.3}),
$g,\dot g$, and $\varepsilon$ are continuous at $t=0$, and we
may assume that $\Psi$ is continuous as well. So that the state
\begin{equation}
\omega=(g,\dot g,\Psi)
\label{2.19}
\end{equation}
is continuous at $t=0$. The metric is $C^{1}$. Such a singularity
is trivially passable.

\subsection{Pulsating black hole}

It is obvious what is the time evolution of the system
considered. The time $\tau/2$ is given by [8]
\begin{equation}
\frac{\tau}{2}=\frac{\pi}{2\sqrt{k}}=\frac{\pi}{2}
\left(\frac{3}{8\pi\kappa\rho(0)}\right)^{1/2}.
\label{2.20}
\end{equation}
We have in terms of $R$:
\begin{equation}
R(-\tau/2)=1,\; \dot R(-\tau/2)=0;\quad  R(0)=0,\;
\dot R(0)=0;\quad  R(\tau/2)=1,\; \dot R(\tau/2)=0;\quad
{\rm and\; so\; on.}
\label{2.21}
\end{equation}
Thus the ball, or the relating black hole pulsates with the
period $\tau$.

It must be emphasized that the case in point is a pulsating
black hole rather than a black hole and a white one that are
periodically interconverted.

\subsection{Escape effect}

Now we should like to take into account the quantum nature
of matter. To do this we consider the motion of a quantum
particle in a gravitational potential well generated by the
dust ball.

From eqs.(\ref{2.3}),(\ref{2.6}),(\ref{2.7}) it follows
\begin{equation}
\frac{d^{2}y}{dt^{2}}=-\frac{\kappa M}{y^{2}},\quad y=aR.
\label{2.22}
\end{equation}
An analogous equation holds [9] for a quantity
\begin{equation}
\eta=aR_{\eta},\quad R_{\eta}<R,
\label{2.23}
\end{equation}
i.e.,
\begin{equation}
\frac{d^{2}\eta}{dt^{2}}=-\frac{\kappa M_{\eta}}{\eta^{2}}
\label{2.24}
\end{equation}
where
\begin{equation}
M_{\eta}=M(\eta/y)^{3}
\label{2.25}
\end{equation}
is the mass of the ball with radius $\eta$. Equations
(\ref{2.24}),(\ref{2.25}) imply that the potential energy of
a particle with mass $m$ is
\begin{equation}
U(\eta)=\left\{
\begin{array}{rcl}
-u+(\gamma/2)\eta^{2}\quad {\rm for}\;\; \eta\le y\\
-\kappa Mm/\eta\quad {\rm for}\;\; \eta\ge y.\\
\end{array}
\right.
\label{2.26}
\end{equation}
From the conditions
\begin{equation}
U\mid_{y+0}=U\mid_{y-0},\quad \frac{dU}{d\eta}\mid_{y+0}=
\frac{dU}{d\eta}\mid_{y-0}
\label{2.27}
\end{equation}
it follows
\begin{equation}
U(\eta)=\frac{\kappa Mm}{2y^{3}}(\eta^{2}-3y^{2})\quad
{\rm for}\;\; \eta\le y.
\label{2.28}
\end{equation}
We find for the average value inside the ball
\begin{equation}
\langle U\rangle=\frac{1}{y^{3}/3}\int_{0}^{y}d\eta\,\eta^{2}
U(\eta)=-\frac{6\kappa Mm}{5y}\cong-\frac{\kappa Mm}{y}.
\label{2.29}
\end{equation}
Thus we may consider the particle in a well of radius
$y$ and depth
\begin{equation}
U_{0}(y)=\frac{\kappa Mm}{y}.
\label{2.30}
\end{equation}
The criterion for the existence of at least one discrete level
is
\begin{equation}
U_{0}y^{2}>\frac{\pi^{2}}{8m}\cong\frac{1}{m},
\label{2.31}
\end{equation}
which is valid both for the nonrelativistic and the
relativistic
case [10]. Thus for
\begin{equation}
y<\frac{1}{\kappa Mm_{max}^{2}},
\quad m_{max}<M,
\label{2.32}
\end{equation}
there is no matter inside the ball. We call this the escape
effect. Needless to say, the effect is quantum.

\subsection{Violation of the constraint equation of
general relativity}

The $0\mu$-components of the Einstein equation give
\begin{equation}
G_{0\mu}=8\pi\kappa T_{0\mu}\Rightarrow \dot R^{2}+k=
\frac{8\pi\kappa}{3}\rho(t)R^{2},
\label{2.33}
\end{equation}
which is the constraint equation for the problem considered.
After the escape, eq.(\ref{2.33}) reduces to
\begin{equation}
\dot R^{2}+k=0,
\label{2.34}
\end{equation}
which contradicts eq.(\ref{2.6}). This result shows once again
that generally only the projected Einstein equation is fulfilled.

\subsection{Naked singularity}

For a naked singularity, eq.(\ref{2.6}) is replaced by [11]
\begin{equation}
\dot R^{2}=\frac{F(r)}{R}+f(r),
\label{2.35}
\end{equation}
so that the principal results do not change.

\section{Big cram: big crunch and big bang}

\subsection{The Robertson-Walker spacetime and the
cosmic-length universe}

We consider a global, or cosmic singularity for the
cosmic-length universe [3,4]---a model based on the
Robertson-Walker spacetime.

The Robertson-Walker metric is of the form
\begin{equation}
g=dt^{2}-R^{2}(t)\left\{\frac{dr^{2}}{1-kr^{2}}+
r^{2}d\theta^{2}+
r^{2}\sin^{2}\theta\,d\varphi^{2}\right\},\quad k=-1,0,1.
\label{3.1}
\end{equation}
We are interested in the closed universe, $k=1$. The projected
Einstein equation gives
\begin{equation}
G_{ij}=8\pi\kappa T_{ij}\Rightarrow 2\ddot RR+\dot R^{2}+k
=-8\pi\kappa pR^{2}.
\label{3.2}
\end{equation}
The equation of motion for matter is
\begin{equation}
T^{0\nu}{}_{;\nu}=0,
\label{3.3}
\end{equation}
or
\begin{equation}
d\varepsilon=-pdV,
\label{3.4}
\end{equation}
where
\begin{equation}
\varepsilon=\rho V,
\label{3.5}
\end{equation}
\begin{equation}
V=2\pi^{2}R^{3}.
\label{3.6}
\end{equation}
Eqs.(\ref{3.2}),(\ref{3.4}) lead to
\begin{equation}
R\dot R^{2}+kR-\frac{8\pi\kappa}{3}\rho R^{3}=L={\rm const}
\label{3.7}
\end{equation}
where $L$ is the cosmic length [4].

\subsection{Equation of state}

The equation of state for matter is
\begin{equation}
p=\gamma\rho,\quad \gamma=\gamma(V),\quad -1\le\gamma\le
\frac{1}{3}.
\label{3.8}
\end{equation}
From eqs.(\ref{3.4}),(\ref{3.8}) it follows
\begin{equation}
\frac{d\rho}{\rho}=-\frac{1+\gamma(V)}{V}dV,
\label{3.9}
\end{equation}
whence
\begin{equation}
\frac{\rho V}{\rho_{0}V_{0}}\exp\left\{\int_{V_{0}}^{V}
\frac{\gamma(V')}{V'}dV'\right\}=1.
\label{3.10}
\end{equation}
With singularity in mind, we put
\begin{equation}
V_{0}\to+0,\quad V\to+0,\quad \gamma(V')=\gamma(0)
\label{3.11}
\end{equation}
and obtain
\begin{equation}
\rho V^{1+\gamma(0)}={\rm const},
\label{3.12}
\end{equation}
which is the equation of state of matter for small $V$.

\subsection{Equation of motion for metric}

From here on we shall use the equation of state of matter
(\ref{3.12}). We have \begin{equation}
\varepsilon=\frac{\beta}{R^{3\gamma(0)}},\quad \beta=
{\rm const}.
\label{3.13}
\end{equation}
From eqs.(\ref{3.7}),(\ref{3.5}),(\ref{3.6}),(\ref{3.13}) we
obtain
\begin{equation}
R\dot R^{2}+kR=\frac{B}{R^{3\gamma(0)}}+L
\label{3.14}
\end{equation}
where
\begin{equation}
B=\frac{4\kappa\beta}{3\pi}.
\label{3.15}
\end{equation}

The metric (\ref{3.1}) depends on
\begin{equation}
q=R^{2}.
\label{3.16}
\end{equation}
We obtain in terms of $q$
\begin{equation}
\varepsilon=\frac{\beta}{q^{3\gamma(0)/2}}
\label{3.17}
\end{equation}
and the equation of motion for the metric:
\begin{equation}
\dot q^{2}+4kq=4q^{1/2}\left[\frac{B}{q^{3\gamma(0)/2}}+
L\right].
\label{3.18}
\end{equation}

\subsection{Singularity}

In the neighborhood of the singularity $q=0$, eq.(\ref{3.18})
reduces to
\begin{equation}
\dot q^{2}=4q^{1/2}\left[\frac{B}{q^{3\gamma(0)/2}}+L\right].
\label{3.19}
\end{equation}
Let $q=0$ correspond to $t=0$.

\subsection{Nonpositive pressure: tempered singularity and
trivial passage}

Let
\begin{equation}
-1\le\gamma(0)\le 0
\label{3.20}
\end{equation}
(the values $-1$ and $0$ relate to the false vacuum and dust
respectively). Then eq.(\ref{3.19}) reduces to
\begin{equation}
\dot q^{2}=4L_{B}q^{1/2}
\label{3.21}
\end{equation}
where
\begin{equation}
L_{B}=\left\{
\begin{array}{rcl}
L\quad {\rm for}\;\; \gamma(0)<0\\
L+B\quad {\rm for}\;\; \gamma(0)=0,\\
\end{array}
\right.
\label{3.22}
\end{equation}
so that
\begin{equation}
\frac{dq}{dt}=\mp 2L^{1/2}_{B}q^{1/4}.
\label{3.23}
\end{equation}
Taking into account eqs.(\ref{2.14})-(\ref{2.18}),
we obtain
\begin{equation}
q=\left(\frac{3}{2}\right)^{4/3}L^{2/3}_{B}t^{4/3},
\label{3.24}
\end{equation}
\begin{equation}
\dot q=2\left(\frac{3}{2}\right)^{1/3}L^{2/3}_{B}t^{1/3},
\label{3.25}
\end{equation}
\begin{equation}
\ddot q=\left(\frac{2}{3}\right)^{2/3}L^{2/3}_{B}t^{-2/3}.
\label{3.26}
\end{equation}
Eq.(\ref{3.17}) results in
\begin{equation}
\varepsilon=\beta\left[\left(\frac{3}{2}\right)^{2}L_{B}\right]
^{\mid\gamma(0)\mid}t^{2\mid\gamma(0)\mid}.
\label{3.27}
\end{equation}

Thus the singularity is tempered and the metric is $C^{1}$.
The passage of the singularity is trivial.

\subsection{Positive pressure: strong singularity and CPT
passage}

Let
\begin{equation}
0<\gamma(0)\le \frac{1}{3}
\label{3.28}
\end{equation}
(the value 1/3 relates to relativistic matter).
From eq.(\ref{3.17}) it follows
\begin{equation}
\lim_{q\to 0}\varepsilon=\infty,
\label{3.29}
\end{equation}
so that the singularity is strong. Eq.(\ref{3.19}) reduces to
\begin{equation}
\dot q^{2}=4Bq^{[1-3\gamma(0)]/2},
\label{3.30}
\end{equation}
whence
\begin{equation}
\frac{dq}{dt}=\mp 2B^{1/2}q^{\alpha}
\label{3.31}
\end{equation}
where
\begin{equation}
\alpha=\frac{1-3\gamma(0)}{4},\qquad 0\le\alpha<\frac{1}{4}.
\label{3.32}
\end{equation}

We obtain for $\alpha>0$:
\begin{equation}
q=w\,t^{1/(1-\alpha)},
\label{3.33}
\end{equation}
\begin{equation}
\dot q=\frac{1}{1-\alpha}w\,t^{\alpha/(1-\alpha)},
\label{3.34}
\end{equation}
\begin{equation}
\ddot q=\frac{\alpha}{(1-\alpha)^{2}}w\,t^{-(1-2\alpha)/
(1-\alpha)},
\label{3.35}
\end{equation}
where
\begin{equation}
w=[2B^{1/2}(1-\alpha)]^{1/(1-\alpha)};
\label{3.36}
\end{equation}
the metric is $C^{1}$.

We have for $\alpha=0$:
\begin{equation}
q=2B^{1/2}\mid t\mid,
\label{3.37}
\end{equation}
\begin{equation}
\dot q=\left\{
\begin{array}{rcl}
-2B^{1/2}\quad {\rm for}\;\; t<0\\
2B^{1/2}\quad {\rm for}\;\; t>0,
\end{array}
\right.
\label{3.38}
\end{equation}
\begin{equation}
\ddot q=4B^{1/2}\delta(t);
\label{3.39}
\end{equation}
the metric is $C$.

Thus, for a positive pressure, the cosmic singularity is
strong and the passage is a $CPT$ one.

\subsection{Quantum matter}

Now let us take into account the quantum nature of matter.
In the case of the closed universe with radius $R$, we
have for the momentum of a quantum particle
\begin{equation}
p\ge\frac{1}{R},
\label{3.40}
\end{equation}
so that on account of
\begin{equation}
E^{2}=m^{2}+p^{2},
\label{3.41}
\end{equation}
we obtain
\begin{equation}
E>\frac{1}{R}=\frac{1}{q^{1/2}}.
\label{3.42}
\end{equation}
Therefore, in view of eq.(\ref{3.17}), if there is some matter
in the vicinity of the big cram,
\begin{equation}
\gamma(0)=\frac{1}{3}
\label{3.43}
\end{equation}
holds.

Thus there are two possibilities for the global singularity:
\begin{equation}
\gamma(0)=-1,\qquad {\rm the\; false\; vacuum};
\label{3.44}
\end{equation}
\begin{equation}
\gamma(0)=\frac{1}{3},\qquad {\rm some\; relativistic\; matter.}
\label{3.44a}
\end{equation}

In the case of the false vacuum, we obtain from
eqs.(\ref{3.22}),(\ref{3.24})-(\ref{3.27}):
\begin{equation}
q=\left(\frac{3}{2}\right)^{4/3}L^{2/3}\,t^{4/3},
\label{3.45}
\end{equation}
\begin{equation}
\dot q=2\left(\frac{3}{2}\right)^{1/3}L^{2/3}\,t^{1/3},
\label{3.45a}
\end{equation}
\begin{equation}
\ddot q=\left(\frac{2}{3}\right)^{2/3}L^{2/3}\,t^{-2/3},
\label{3.45b}
\end{equation}
and
\begin{equation}
\varepsilon=\beta\left(\frac{3}{2}\right)^{2}L\,t^{2};
\label{3.46}
\end{equation}
the singularity is tempered, the metric is $C^{1}$, and the
passage is trivial.

In the case of relativistic matter, we obtain from
eqs.(\ref{3.37})-(\ref{3.39}),(\ref{3.17}),(\ref{3.15}):
\begin{equation}
q=2B^{1/2}\mid t\mid,
\label{3.47}
\end{equation}
\begin{equation}
\dot q=2B^{1/2}[-\theta(-t)+\theta(t)],
\label{3.47a}
\end{equation}
\begin{equation}
\ddot q=4B^{1/2}\delta(t),
\label{3.47b}
\end{equation}
and
\begin{equation}
\varepsilon=\frac{(3\pi)^{1/4}\beta^{3/4}}{2\kappa^{1/4}}
\frac{1}{\mid t\mid^{1/2}};
\label{3.48}
\end{equation}
the singularity is strong, the metric is $C$, and the passage
is a $CPT$ one.

\subsection{The oscillating universe}

The passableness for the cosmic singularity implies the
oscillating universe (for some discussion, see [1]). We have
for the universe:
\begin{equation}
{\rm ...\quad  expansion,\quad contraction,
\quad big\; cram\;(big\;
crunch\;followed\;by\;big\;bang),\quad
expansion\quad ...}
\label{3.49}
\end{equation}

\end{document}